\documentclass[twocolumn,amsmath,amssymb]{revtex4}
\usepackage{amsfonts}
\usepackage[usenames]{color}

\newcommand{\be}{\begin{equation}}
\newcommand{\ee}{\end{equation}}
\newcommand{\bea}{\begin{eqnarray}}
\newcommand{\eea}{\end{eqnarray}}

\begin{document}
\title{Graviton-photon conversions in Euler-Heisenberg
nonlinear electrodynamics}
\author{Jai-chan Hwang${}^{1}$, Hyerim Noh${}^{2}$}
\address{${}^{1}$Particle Theory and Cosmology Group,
         Center for Theoretical Physics of the Universe,
         Institute for Basic Science (IBS), Daejeon, 34126, Republic of Korea
         \\
         ${}^{2}$Theoretical Astrophysics Group, Korea Astronomy and Space Science Institute, Daejeon, Republic of Korea
         }


\begin{abstract}

We study graviton-photon conversions in an environment of the uniform and constant magnetic field considering Euler-K\"ockel-Heisenberg-type nonlinear corrections in electrodynamics. We take the transverse-tracefree gauge for gravitons and employ both the field potential and the electric and magnetic (EM) fields for photons. The nonlinear correction causes parity violating (chiral) graviton propagation equations, which depend on the two treatments for photons. Thus, the medium becomes effectively birefringent for two polarizations of gravitational waves similar to the photon birefringence a characteristic of the nonlinear correction. In the presence of gravity, due to the nontrivial relation between the potential and EM fields, it is important to present the result using the EM fields.

\end{abstract}

\maketitle

%
%
%
\section{Introduction}

The graviton-photon conversions mediated by external magnetic field, known as Gertsenshtein and its inverse mechanism, are processes naturally arising in classical Maxwell's equations in the curved spacetime of Einstein's gravity \cite{Gertsenshtein-1961, Lupanov-1967, Boccaletti-1970, Zeldovich-1973, Zeldovich-Novikov-1983, Raffelt-1988}. The mechanism can be used to detect gravitational waves by observing the converted photons in the presence of external magnetic field in celestial objects or in the Laboratory \cite{Aggarwal-2021, Domcke-2021, Vagnozzi-2022, Liu-2023, Ramazanov-2023}.

In the graviton-photon conversions, the presence of gravitons in the energy-momentum tensor, which works as a source of the graviton, is often ignored. The ignored term, however, is of the same order as the other terms. Recently, by including its effect, we show that the effect causes a negative effective mass-squared (tachyonic instability) term in the graviton propagation equation, and showed that the effect depends on whether we use the potential or EM fields \cite{Hwang-Noh-2023-GP}.

The transverse-tracefree (TT) gauge is often adopted for gravitons and the external (or background) magnetic field is usually assumed to be uniform and constant with its gravity often ignored in the analysis. These are assumptions. We can relax these assumptions on the background magnetic field and take other gauges for gravitons. For example, in the case of gravitational wave detection using the electromagnetic method, only the graviton to photon conversion is considered with the Fermi normal coordinate preferred for the detector instead of the TT gauge \cite{Fortini-1982, Baroni-1984, 
Flores-1986, Marzlin-1994, Licht-2004, Rakhmanov-2014}.

In a strong magnetic field, the quantum correction can cause nonlinear corrections in electrodynamics, interpreted as the vacuum polarization. Euler-K\"ockel-Heisenberg correction is well known in quantum electrodynamics (QED) \cite{Euler-Kockel-1935, Euler-1936, Heisenberg-Euler-1936}. There are other similarly motivated nonlinear corrections \cite{Born-1934, Born-Infeld-1934, Sorokin-2022}. The nonlinear correction is often included in the graviton-photon conversion \cite{Raffelt-1988, Dolgov-Ejlli-2012, Dolgov-Ejlli-2013, Ejlli-Dolgov-2014, Ramazanov-2023} and axion-photon conversion \cite{Raffelt-1988}. Previous studies exclusively used the potential for photons.

In electrodynamics, we have two options of using the electric and magnetic (EM) fields or using the field potential. As a natural solution of the homogeneous Maxwell's equation $\nabla_{[a} F_{bc]} = 0$, the potential is introduced as $F_{ab} \equiv \nabla_a A_b - \nabla_b A_a$; the covariant derivatives become ordinary derivatives and these are valid in curved spacetime of gravity. In the absence of gravity the relation between the potential and EM fields is trivial and well known \cite{Jackson-1975}: ${\bf E} = - \nabla \phi - { 1\over c} \dot {\bf A}$ and ${\bf B} = \nabla \times {\bf A}$ with $\phi = - A_0$. Thus, we can easily recover the EM fields from the potential.

In curved spacetime the relation becomes complicated. The EM fields, being measurable quantities by a given observer, are defined from the field-strength tensor in a sophisticated manner \cite{Moller-1952, Lichnerowicz-1967, Ellis-1973}. Even in the flat spacetime the EM fields are defined by decomposing the field-strength tensor using a four-vector associated with the observer. The above EM fields are the quantities measured by an Eulerian observer associated with the normal four-vector \cite{Smarr-York-1978, Wilson-Mathews-2003, Gourgoulhon-2012} which can be chosen as $n_a = (-1, 0, 0, 0)$ in flat spacetime.

The relativistic gravity causes nontrivial modifications in the relation between the potential and EM fields \cite{Thorne-MacDonald-1982, Hwang-Noh-2023-EM-NL}. Using the potential, as the covariant derivatives can be replaced by partial derivatives, with $F_{ab} = \partial_a A_b - \partial_b A_a$, the field-strength tensor $F_{ab}$ is free from the metric and the homogeneous Maxwell's equation, with $\partial_{[a} F_{bc]} = 0$, is identically valid. Whereas, the EM fields are introduced by decomposing $F_{ab}$ using the observer's four-vector $u_a$ \cite{Moller-1952, Lichnerowicz-1967, Ellis-1973}. In this way, using the EM fields, for any observer $F_{ab}$ with two covariant indices and the homogeneous equation involve the metric, see \cite{EM-definition, Hwang-Noh-2023-EM-NL}. 
As the EM fields are gauge-invariant and more directly related to measured quantities compared to the potential, in a curved spacetime, it is essential to analyze using the EM fields or transform the result using the EM fields.

Previously, we studied the graviton-photon conversion mechanism in classical electrodynamics \cite{Hwang-Noh-2023-GP}. Here, we extend the study to include the nonlinear corrections. We study the mechanism using the two methods to handle photons. We show that the graviton propagation equations show parity violating (chiral) terms appearing in the nonlinear corrections, and the behaviors depend on the two methods.

In Sec.\ \ref{sec:NL} we summarize a non-linear electrodynamics made of general function of two invariants. In Sec.\ \ref{sec:GP} we study the graviton-photon conversion including the leading order nonlinear corrections using the EM fields and potential for photons. In Sec.\ \ref{sec:birefringence} we present the well known photon birefringence using the two methods. In Sec.\ \ref{sec:Discussion} we discuss the results.

%
%
%
\section{Nonlinear electrodynamics}
                               \label{sec:NL}

We consider a general action of the EM part \cite{Peres-1961}
\bea
   & & {\cal L} = \sqrt{-g} [
       L (I, J) + {1 \over c} J^a A_a ],
   \nonumber \\
   & & I \equiv {1 \over 4} F_{ab} F^{ab}
       = {1 \over 2} ( B^2 - E^2 ),
   \nonumber \\
   & &
       J \equiv {1 \over 4} F_{ab} F^{*ab}
       = - E^a B_a,
   \label{L}
\eea
where $L$ is a general algebraic function of $I$ and $J$ which are unique Lorentz- and gauge-invariant terms without derivatives; $E^2 \equiv E^a E_a$, etc. Variations with respect to $g_{ab}$ and $A_a$, respectively, give
\bea
   & & T_{ab} = - T_{ab}^{\rm EM} L_{,I}
       + g_{ab} ( L - I L_{,I} - J L_{,J} ),
   \label{Tab}\\
   & & H^{ab}_{\;\;\;\;;b} = {1 \over c} J^a,
   \label{ME-Hab}
\eea
where
\bea
   & & T_{ab}^{\rm EM} = F_{ac} F_b^{\;\;c}
       - {1 \over 4} g_{ab} F^{cd} F_{cd},
   \\
   & & H_{ab} \equiv - F_{ab} L_{,I} - F^*_{ab} L_{,J},
   \label{Hab-def}
\eea
and $L_{,I} \equiv \partial L/\partial I$, etc; $T_{ab}^{\rm EM}$ is tracefree, but not for $T_{ab}$ in Eq.\ (\ref{Tab}). We took the Heaviside-Lorentz unit; translation to Gaussian unit will be explained below. The other Maxwell's equation is
\bea
   \eta^{abcd} F_{bc,d} = 0, \quad
   {\rm or} \quad
   F^{*ab}_{\;\;\;\;\;\;;b} = 0.
   \label{ME-Fab}
\eea
Thus, the nonlinear corrections can be interpreted as the effective medium property or the effective four-current in the conventional Maxwell's equations.

Using a time-like four-vector $u_a$, EM fields are introduced using the covariant decomposition as \cite{Ellis-1973},
\bea
   & & F_{ab} \equiv u_a E_b - u_b E_a
       - \eta_{abcd} u^c B^d,
   \nonumber \\
   & &
       H_{ab} \equiv u_a D_b - u_b D_a
       - \eta_{abcd} u^c H^d,
   \nonumber \\
   & & F^*_{ab} = u_a B_b - u_b B_a
       + \eta_{abcd} u^c E^d,
   \nonumber \\
   & &
       H^*_{ab} = u_a H_b - u_b H_a
       + \eta_{abcd} u^c D^d,
   \label{Hab-cov}
\eea
with $E_a u^a \equiv 0$ etc., thus, $E_a \equiv F_{ab} u^b$, $B_a \equiv F^*_{ab} u^b$, $D_a \equiv H_{ab} u^b$, and $H_a \equiv H^*_{ab} u^b$. Equation (\ref{Hab-def}) gives
\bea
   & & D_a = - E_a L_{,I} - B_a L_{,J} \equiv E_a + P_a,
   \nonumber \\
   & & H_a = - B_a L_{,I} + E_a L_{,J} \equiv B_a - M_a,
\eea
where $P_a$ and $M_a$ are effective polarization and magnetization, respectively. Therefore, nonlinear corrections in Maxwell's equations cause modification of the effective vacuum structure, and such terms arising from quantum correction is called the vacuum polarization.

The four-current can be decomposed to the charge and current densities using the same $u_a$
\bea
   \varrho_{\rm em} c = - J^a u_a, \quad
       j^{a} = h^a_b J^b,
\eea
where $h_{ab} \equiv g_{ab} + u_a u_b$ is the spatial projection tensor; this differs from the perturbed metric in next sections. Equations (\ref{ME-Hab}) and (\ref{ME-Fab}) projecting to $u_a$ and $h_a^c$, respectively, give Maxwell's equations in terms of the EM fields \cite{Ellis-1973, Hwang-Noh-2023-Axion-EM}
\bea
   & & D^a_{\;\; ;b} h^b_a
       = \varrho_{\rm em}
       - 2 \omega^a H_a,
   \label{Maxwell-cov-1} \\
   & & h^a_b \widetilde {\dot D}{}^b
       = \left( \eta^a_{\;\;bcd} u^d \omega^c
       + \sigma^a_{\;\;b}
       - {2 \over 3} \delta^a_b \theta \right) D^b
   \nonumber \\
   & & \qquad
       + \eta^{abcd} u_d
       \left( a_b H_c - H_{b;c} \right)
       - {1 \over c} j^{a},
   \label{Maxwell-cov-2} \\
   & & B^a_{\;\; ;b} h^b_a
       = 2 \omega^a E_a,
   \label{Maxwell-cov-3} \\
   & & h^a_b \widetilde {\dot B}{}^b
       = \left( \eta^a_{\;\;bcd} u^d \omega^c
       + \sigma^a_{\;\;b}
       - {2 \over 3} \delta^a_b \theta \right) B^b
   \nonumber \\
   & & \qquad
       - \eta^{abcd} u_d
       \left( a_b E_c - E_{b;c} \right),
   \label{Maxwell-cov-4}
\eea
where $\widetilde {\dot E}{}^a \equiv E^a_{\;\; ;b} u^b$; $\omega_a$, $\sigma_{ab}$, $a_a$, and $\theta$ are the vorticity vector, shear tensor, acceleration vector, and expansion scalar, respectively, of the $u_a$-flow \cite{Ellis-1971}. In this way, the nonlinear corrections in Maxwell's equation in Eq.\ (\ref{ME-Hab}) can be interpreted as medium properties in the Maxwell's equations in general relativity \cite{Ellis-1973}.

\subsection{Leading order nonlinear electrodynamics}

In the following we consider the leading order nonlinear correction with
\bea
   L = - I + \varrho \Big( 2 \alpha I^2
       + {7 \over 2} \beta J^2 \Big).
   \label{l-EH}
\eea
Two constants $\alpha$ and $\beta$ are introduced to trace the two terms; $\alpha = \beta = 1$ in the Euler-K\"ockel-Heisenberg case based on QED \cite{Euler-Kockel-1935, Euler-1936, Heisenberg-Euler-1936} and other values in other cases \cite{Born-1934, Born-Infeld-1934, Sorokin-2022}; $\varrho$ is a constant with the dimension of $B^{-2}$.

In QED, we have $\varrho \equiv (\bar \alpha/ 45 \pi) ( e \hbar/ m_e^2 c^3 )^2$ \cite{Brezin-1971} with $\bar \alpha$ the fine-structure constant. In this form of $\varrho$ the Lagrangian is invariant under a change to Gaussian unit, except for an overall factor of $1/(4 \pi)$; under the unit change, we have $(F_{ab},E_a, B_a, A_a) \rightarrow (F_{ab},E_a, B_a, A_a) / \sqrt{4 \pi}$ and and $(J_a, P_a, M_a, j_a, \varrho, e) \rightarrow \sqrt{4 \pi} (J_a, P_a, M_a, j_a, \varrho, e)$. Transforming to Gaussian unit, $\varrho I$ and $\varrho J$ are invariant, photon conversion equations remain the same, and graviton conversion equations have $1/(4 \pi)$ factors appearing.

Equations become
\bea
   & & ( \sqrt{-g} H^{ab} )_{,b} = {1 \over c} \sqrt{-g} J^a,
   \label{Fab-eq} \\
   & & T_{ab} = T_{ab}^{\rm EM}
       - \alpha \varrho F^2 ( F_{ac} F_b^{\;\;c}
       - {1 \over 8} g_{ab} F^2 )
   \nonumber \\
   & & \qquad
       - {7 \over 32} \beta \varrho g_{ab} (F F^*)^2,
   \label{Tab-EH} \\
   & & H_{ab} = ( 1 - \alpha \varrho F^2 ) F_{ab}
       - {7 \over 4} \beta \varrho F F^* F^*_{ab},
   \nonumber \\
   & & D_a = ( 1 - \alpha \varrho F^2 ) E_a
       - {7 \over 4} \beta \varrho F F^* B_a,
   \nonumber \\
   & &
       H_a = ( 1 - \alpha \varrho F^2 ) B_a
       + {7 \over 4} \beta \varrho F F^* E_a,
   \label{Hab-EH}
\eea
where $F^2 \equiv F_{ab} F^{ab}$ and $F F^* \equiv F_{ab} F^{*ab}$. All results in this section are spacetime covariant. In the following, in case we need clear distinction, we will indicate covariant quantities using an overtilde like $\widetilde B_a$.

%
%
%
\section{Graviton-photon conversions}
                              \label{sec:GP}

In the graviton-photon conversions we {\it assume} a uniform and constant magnetic field as an environment and consider the photons and gravitational waves as the linear order perturbations \cite{Hwang-Noh-2023-GP}. Perturbed metric in Minkowski background is
\bea
   g_{ab} = \eta_{ab} + h_{ab}.
\eea

The EM fields are defined using the covariant decomposition in Eq.\ (\ref{Hab-cov}) based on the normal frame four-vector, see Eq.\ (10) in \cite{EM-definition}. We introduce $\widetilde E_i \equiv E_i$ where the index of $E_i$ is associated with $\delta_{ij}$ and its inverse. We decompose the EM fields to the background fields and generated photons, and {\it ignore} the background $E_i$, thus $B_i \rightarrow B_i + b_i$ and $E_i \rightarrow e_i$; $b_i$ and $e_i$ are perturbed order EM fields, i.e., photons. Under TT gauge, with $h^j_{i,j} \equiv 0 \equiv h^j_j$ and $h_{00} \equiv 0 \equiv h_{0i}$, we have \cite{Hwang-Noh-2023-GP}
\bea
   F_{0i} = - e_i, \quad
       F_{ij} = \eta_{ijk} ( B^k + b^k - h^{k\ell} B_\ell ).
   \label{Fab-EB}
\eea
Notice the gravitational waves appearing in $F_{ij}$.

In terms of the field potential for photons, defined as $F_{ab} \equiv \partial_a A_b - \partial_b A_a$, we have
\bea
   F_{0i} = \partial_0 A_i - \partial_i A_0, \quad
       F_{ij} = \eta_{ijk} B^k
       + \partial_i A_j - \partial_j A_i,
   \label{Fab-EB-A}
\eea
where we regard $A_a$ as photons only; the background magnetic field is represented by $B_i$. In terms of potential, apparently, the above $F_{ab}$ with two covariant (lower) indices does not involve the metric perturbation. We set $\widetilde A_i \equiv A_i$ where the index of $A_i$ is associated with $\delta_{ij}$ and its inverse. Indices of $B_i$ for the background magnetic field and $b_i$, $e_i$, and $A_i$ for photons are associated with $\delta_{ij}$.

\subsection{Graviton conversion}
                                        \label{sec:graviton}

Einstein's equation in transverse-tracefree (TT) gauge condition gives \cite{Hwang-Noh-2023-GP}
\bea
   ( \partial_0^2 - \Delta ) h_{ij}
       = {16 \pi G \over c^4} \delta T_{ij}^{\rm TT},
   \label{GW-eq-ij}
\eea
where $\delta T_{ij}^{\rm TT}$ is the TT projection of perturbed part of $T_{ij}$, see \cite{Maggiore-2007, Hwang-Noh-2023-GP}. In the graviton-photon conversion we often ignore the gravity of the background magnetic field affecting the background geometry. This is justified in practice. The conversion rate is directly related to the gravitational strength of the background magnetic field and is extremely small in conceivable astrophysical and Laboratory situations: see Eqs.\ (20) and (75) in \cite{Hwang-Noh-2023-GP}.

\subsubsection{In terms of EM fields}

From Eq.\ (\ref{Tab-EH}), we have
\begin{widetext}
\bea
   & & T_{ij} = - B_i B_j + {1 \over 2} \delta_{ij} B^2
       - B_i b_j - B_j b_i
       + \delta_{ij} {\bf B} \cdot {\bf b}
       + {1 \over 2} ( h_{ij} B^2
       - \delta_{ij} h^{k\ell} B_k B_\ell )
       + 2 \alpha \varrho \Big[
       B^2 \Big( B_i B_j - {3 \over 4} \delta_{ij} B^2 \Big)
   \nonumber \\
   & & \qquad
       + (B_i b_j + B_j b_i ) B^2
       + {\bf B} \cdot {\bf b}
       ( 2 B_i B_j - 3 \delta_{ij} B^2 )
       - {3 \over 4} h_{ij} B^4
       + \Big( {3 \over 2} B^2 \delta_{ij} - B_i B_j \Big)
       h^{k\ell} B_k B_\ell \Big].
\eea
To the linear order, $\beta$-term disappears in $T_{ij}$. By applying the TT operation in Eq.\ (5) of \cite{Hwang-Noh-2023-GP}, we have
\bea
   & & \delta T^{\rm TT}_{ij}
       = - B_i b_j - B_j b_i
       + \delta_{ij} ( {\bf B} \cdot {\bf b}
       - B^k \hat n_k b^\ell \hat n_\ell )
       + 2 \hat n_{(i} ( B_{j)} b^k \hat n_k
       + b_{j)} B^k \hat n_k )
       - \hat n_i \hat n_j ( {\bf B} \cdot {\bf b}
       + B^k \hat n_k b^\ell \hat n_\ell )
       + {1 \over 2} h_{ij} B^2
   \nonumber \\
   & & \qquad
       + 2 \alpha \varrho \Big\{
       ( B_i b_j + B_j b_i ) B^2
       + {\bf B} \cdot {\bf b}
       ( 2 B_i B_j - 4 B_{(i} \hat n_{j)} B^k \hat n_k )
       - 2 B^2 \hat n_{(i} ( B_{j)} b^k \hat n_k
       + b_{j)} B^k \hat n_k )
   \nonumber \\
   & & \qquad
       + \delta_{ij} \big[ - 2 B^2 {\bf B} \cdot {\bf b}
       + ( B^k \hat n_k )^2 {\bf B} \cdot {\bf b}
       + B^2 B^k \hat n_k b^\ell \hat n_\ell \big]
       + \hat n_i \hat n_j \big[ 2 B^2 {\bf B} \cdot {\bf b}
       + ( B^k \hat n_k )^2 {\bf B} \cdot {\bf b}
       + B^2 B^k \hat n_k b^\ell \hat n_\ell \big]
   \nonumber \\
   & & \qquad
       - {3 \over 4} h_{ij} B^4
       + h^{k\ell} B_k B_\ell \Big[
       - B_i B_j + {1 \over 2} \delta_{ij}
       \big[ B^2 - (B^m \hat n_m)^2 \big]
       + 2 \hat n_{(i} B_{j)} \hat n_m B^m
       - {1 \over 2} \hat n_i \hat n_j
       \big[ B^2 + (B^m \hat n_m)^2 \big]
       \Big] \Big\},
\eea
where $A_{(ij)} \equiv {1 \over 2} (A_{ij} + A_{ji})$.
We align the $z$-axis to the propagating direction of a plane gravitational wave. Without losing generality (using the cylindrical symmetry), we align the background magnetic field in the $y$-$z$ plane with $B_2 = B \sin{\theta}$, $B_3 = B \cos{\theta}$, and $B_1 = 0$. Then, we have
\bea
   & & \delta T_{11}^{\rm TT}
       = - \delta T_{22}^{\rm TT}
       = B_2 b_2 + {1 \over 2} B^2 h_+
       - \alpha \varrho
       \Big[ 2 ( 2 B_2^2 + B_3^2 ) B_2 b_2 + 2 B_2^2 B_3 b_3
       + \Big( {3 \over 2} B^4 + B_2^4 \Big) h_+ \Big],
   \nonumber \\
   & & \delta T_{12}^{\rm TT}
       = - B_2 b_1 + {1 \over 2} B^2 h_\times
       + \alpha \varrho
       \Big[ 2 ( B_2^2 + B_3^2 ) B_2 b_1
       - {3 \over 2} B^4 h_\times \Big],
\eea
and zeros otherwise. Using the two polarizations of gravitons introduced as $h_+ \equiv h_{11} = - h_{22} $ and $h_\times \equiv h_{12} = h_{21}$, Eq.\ (\ref{GW-eq-ij}) gives
\bea
   & & \hskip -.8cm
       \Big\{ \partial_0^2 - \partial_z^2
       - {8 \pi G \over c^4} B^2 [ 1
       - \alpha \varrho B^2 ( 3 + 2 \sin^4{\theta} ) ]
       \Big\} h_{+}
       = {16 \pi G \over c^4} B \sin{\theta}
       \Big\{ b_2
       - 2 \alpha \varrho B^2 [ b_2 ( 1 + \sin^2{\theta} )
       + b_3 \sin{\theta} \cos{\theta} ] \Big\},
   \label{GW-B-1} \\
   & & \hskip -.8cm
       \Big[ \partial_0^2 - \partial_z^2
       - {8 \pi G \over c^4} B^2 ( 1 - 3 \alpha \varrho B^2 )
       \Big] h_{\times}
       = - {16 \pi G \over c^4} B \sin{\theta}
       ( 1 - 2 \alpha \varrho B^2 ) b_1.
   \label{GW-B-2}
\eea
Thus, no graviton conversion occurs along parallel (i.e., along $z$-axis) components of the background magnetic field (i.e., for $\theta = 0, \pi$). Due to the nonlinear effect, the impinging photons along $z$-axis (i.e., $b_3$) contribute to the graviton conversion. The $\alpha$-term causes parity breaking (chiral) effect in the gravitational wave equations for $\theta \neq 0, \pi$. Thus, the $\alpha$-term causes the medium to be effectively birefringent, by making the two polarizations of gravitational waves propagating differently.

Notice the presence of tachyonic instability (or negative effective mass-squared) terms in the graviton part. These arise from the gravitons present in the energy-momentum tensor in Eq.\ (\ref{Tab-EH}). Although we name it a tachyonic instability term, the instability is not realized in our approximation where we ignore the gravity of the background order magnetic field, see \cite{Hwang-Noh-2023-GP}. In Fourier space with $h_{+,\times} \propto e^{i(\omega t - {\bf k} \cdot {\bf x})}$, ignoring the photon source terms, the left-hand sides of Eqs.\ (\ref{GW-B-1}) and (\ref{GW-B-2}) give $\omega^2/c^2 = k^2 - k_B^2$ with
\bea
   k^+_B = {\sqrt{8 \pi G} \over c^2} B
       \Big[ 1
       - {1 \over 2} \alpha \varrho B^2 ( 3 + 2 \sin^4{\theta} ) \Big], \quad
       k^\times_B = {\sqrt{8 \pi G} \over c^2} B
       \Big( 1 - {3 \over 2} \alpha \varrho B^2 \Big).
   \label{k-EM}
\eea
For $k < k_B$ we have exponential instability, $h_{+,\times} \propto e^{\pm \omega t}$ with real $\omega$, but this is forbidden as the condition $k < k_B$ implies the gravity of the background magnetic field no longer negligible, thus violating the basic assumption of the graviton-photon conversion \cite{Hwang-Noh-2023-GP}. The $\alpha$-correction term tends to reduce the effect.

\subsubsection{In terms of the potential}

For the background EM fields and photons, we have Eq.\ (\ref{Fab-EB-A}). From Eq.\ (\ref{Tab-EH}), we have
\bea
   & & \delta T_{ij}
       = - 2 B_{(i} \eta_{j)k\ell} \partial^k A^\ell
       + \delta_{ij} B^k \eta_{k\ell m} \partial^\ell A^m
       + {1 \over 2} ( h_{ij} B^2
       + \delta_{ij} h^{k\ell} B_k B_\ell )
       - 2 h^k_{(i} B_{j)} B_k
       + \alpha \varrho \Big[
       4 B^2 B_{(i} \eta_{j)k\ell} \partial^k A^\ell
   \nonumber \\
   & & \qquad
       + 2 B_k \eta^{k\ell m} \partial_\ell A_m
       ( 2 B_i B_j - 3 B^2 \delta_{ij} )
       - {3 \over 2} h_{ij} B^4
       + ( 2 B_i B_j - 3 \delta_{ij} B^2 ) h^{k\ell} B_k B_\ell
       + 4 B^2 h_{(i}^k B_{j)} B_k \Big].
\eea
The $\beta$-term disappear in $T_{ij}$. By applying the TT operation, we have
\bea
   & & \delta T^{\rm TT}_{ij}
       = - 2 B_{(i} \eta_{j)k\ell} \partial^k A^\ell
       + ( \delta_{ij} - \hat n_i \hat n_j )
       B_k \eta^{k\ell m} \partial_\ell A_m
       - ( \delta_{ij} + \hat n_i \hat n_j ) \hat n_k B^k
       \hat n_\ell \eta^{\ell mn} \partial_m A_n
       + 2 \hat n_{(i} B_{j)} \hat n_k \eta^{k\ell m}
       \partial_\ell A_m
   \nonumber \\
   & & \qquad
       + 2 \hat n_k B^k \hat n_{(i} \eta_{j) \ell m}
       \partial^\ell A^m
       + {1 \over 2} h_{ij} B^2
       + ( \delta_{ij} - \hat n_i \hat n_j ) h^{k\ell} B_k B_\ell
       - 2 h^k_{(i} B_{j)} B_k
       + 2 \hat n_{(i} h_{j)}^k B_k \hat n_\ell B^\ell
   \nonumber \\
   & & \qquad
       + 2 \alpha \varrho \Big\{
       2 B^2 B_{(i} \eta_{j)k\ell} \partial^k A^\ell
       - 2 B^2 \hat n_k B^k \hat n_{(i} \eta_{j)\ell m}
       \partial^\ell A^m
       + \eta^{k\ell m} \partial_\ell A_m \Big[
       2 B_i B_j B_k
   \nonumber \\
   & & \qquad
       + (\delta_{ij} + \hat n_i \hat n_j )
       \big[ B^2 \hat n_k \hat n_\ell B^\ell
       + B_k (B^\ell \hat n_\ell)^2 \big]
       - 2 (\delta_{ij} - \hat n_i \hat n_j ) B^2 B_k
       - 2 B^2 \hat n_{(i} B_{j)} \hat n_k
       - 4 \hat n_{(i} B_{j)} B_k \hat n_\ell B^\ell \Big]
   \nonumber \\
   & & \qquad
       - {3 \over 4} h_{ij} B^4
       + h^{k\ell} B_k B_\ell \Big[ B_i B_j - {3 \over 2} \delta_{ij} B^2
       + {1 \over 2} ( \delta_{ij} + \hat n_i \hat n_j )
       (\hat n_m B^m)^2
       + {3 \over 2} \hat n_i \hat n_j B^2
       - 2 \hat n_m B^m \hat n_{(i} B_{j)} \Big]
   \nonumber \\
   & & \qquad
       + 2 B^2 h^k_{(i} B_{j)} B_k
       - 2 B^2 \hat n_{(i} h_{j)}^k B_k \hat n_\ell B^\ell
       \Big\}.
\eea

We align the gravitational wave propagation in $z$-axis and the background magnetic field in $y$-$z$ plane. We have
\bea
   & & \delta T_{11}^{\rm TT}
       = - \delta T_{22}^{\rm TT}
       = B_2 ( \partial_3 A_1 - \partial_1 A_3 )
       + {1 \over 2} ( B_3^2 - B_2^2 ) h_+
   \nonumber \\
   & & \qquad
       - \alpha \varrho \Big\{
       2 (2 B_2^2 + B_3^2 ) B_2
       ( \partial_3 A_1 - \partial_1 A_3 )
       + 2 B_2^2 B_3 (\partial_1 A_2 - \partial_2 A_1 )
       + \Big[ {3 \over 2} B^4
       - ( 3 B_2^2 + 2 B_3^2 ) B_2^2 \Big] h_+ \Big\},
   \nonumber \\
   & & \delta T_{12}^{\rm TT}
       = - B_2 ( \partial_2 A_3 - \partial_3 A_2 )
       + {1 \over 2} ( B_3^2 - B_2^2 ) h_\times
       + \alpha \varrho \Big[
       2 ( B_2^2 + B_3^2 ) B_2
       (\partial_2 A_3 - \partial_3 A_2 )
       + B^2 \Big( - {3 \over 2} B^2 + 2 B_2^2 \Big)
       h_\times \Big].
\eea
Thus, Eq.\ (\ref{GW-eq-ij}) gives
\bea
   & & \Big\{ \partial_0^2 - \partial_z^2
       - {8 \pi G \over c^4} B^2 \Big[ 1 - 2 \sin^2{\theta}
       - \alpha \varrho B^2 [ 3 - 2 ( 2 + \sin^2{\theta} )
       \sin^2{\theta} ] \Big]
       \Big\} h_{+}
   \nonumber \\
   & & \qquad
       = {16 \pi G \over c^4} B \sin{\theta} \Big\{
       \partial_3 A_1 - \partial_1 A_3
       - \alpha \varrho B^2 [
       2 ( 1 + \sin^2{\theta} )
       ( \partial_3 A_1 - \partial_1 A_3 )
       + 2 \sin{\theta} \cos{\theta}
       ( \partial_1 A_2 - \partial_2 A_1 ) ] \Big\},
   \label{GW-A-1} \\
   & & \Big\{ \partial_0^2 - \partial_z^2
       - {8 \pi G \over c^4} B^2 [ 1 - 2 \sin^2{\theta}
       - \alpha \varrho B^2 ( 3 - 4 \sin^2{\theta} ) ]
       \Big\} h_{\times}
       = - {16 \pi G \over c^4} B \sin{\theta} ( 1
       - 2 \alpha \varrho B^2 )
       ( \partial_2 A_3 - \partial_3 A_2 ).
   \label{GW-A-2}
\eea
Thus, no graviton conversion occurs along parallel components of the background magnetic field (i.e., for $\theta = 0, \pi$). The $\alpha$-term causes parity breaking effect in the gravitational wave equations for $\theta \neq 0, \pi$.

Notice that depending on using either EM fields or the potential, the forms of gravitational wave equations in the left-hand sides of in Eqs.\ (\ref{GW-A-1}) and (\ref{GW-A-2}) differ from Eqs.\ (\ref{GW-B-1}) and (\ref{GW-B-2}) which is the case even for $\alpha = 0$ \cite{Hwang-Noh-2023-GP}. Furthermore, even for $\alpha = 0$, while the negative mass-squared terms in Eqs.\ (\ref{GW-B-1}) and (\ref{GW-B-2}) are independent of $\theta$, in Eqs.\ (\ref{GW-A-1}) and (\ref{GW-A-2}) the terms depend on $\theta$ and become positive for ${\pi \over 4} < \theta < {3 \pi \over 4}$.

Such differences are expected because the relation between the EM fields and potential depends on the metric (gravity) in a non-trivial way \cite{Thorne-MacDonald-1982, Hwang-Noh-2023-EM-NL}. Using
\bea
   b^i = \eta^{ijk} \partial_j A_k + h^i_j B^j, \quad
       e_i = - A_{i,0},
   \label{b-A}
\eea
we can show that Eqs.\ (\ref{GW-A-1}) and (\ref{GW-A-2}) are consistent with Eqs.\ (\ref{GW-B-1}) and (\ref{GW-B-2}). For the relation between EM fields and the potential in a general curved spacetime, see Eq.\ (3.5) in \cite{Thorne-MacDonald-1982} and Eq.\ (89) in \cite{Hwang-Noh-2023-EM-NL}.


Following a similar analysis as above Eq.\ (\ref{k-EM}), we have
\bea
   & & k^+_B = {\sqrt{8 \pi G} \over c^2} B
       \Big\{ \sqrt{1 - 2 \sin^2{\theta}}
       - {\alpha \varrho B^2 \over 2 \sqrt{1 - 2 \sin^2{\theta}}}
       [ 3 - 2 ( 2 + \sin^2{\theta} )
       \sin^2{\theta} ] \Big\},
   \nonumber \\
   & & k^\times_B = {\sqrt{8 \pi G} \over c^2} B
       \Big[ \sqrt{1 - 2 \sin^2{\theta}}
       - {\alpha \varrho B^2 \over 2 \sqrt{1 - 2 \sin^2{\theta}}}
       ( 3 - 4 \sin^2{\theta} ) \Big].
   \label{k-A}
\eea
Compared with the case using the EM fields in Eq.\ (\ref{k-EM}), using the potential the tachyonic instability terms behave as effective mass depending on $\theta$. Considering the gauge dependent nature of the potential, and the observer related nature of the EM fields, although the two sets are consistent, it is important to present and interpret results based on the EM fields.

\subsection{Photon conversion}
                                        \label{sec:photon}

\subsubsection{In terms of EM fields}

From Eq.\ (\ref{Hab-EH}), using Eq.\ (\ref{Hab-cov}), we have
\bea
   F_{0i} = - E_i, \quad
       F_{ij} = \eta_{ijk}
       ( \delta^{k\ell} - h^{k\ell} ) B_\ell, \quad
       F^*_{0i} = - B_i, \quad
       F^*_{ij} = - \eta_{ijk}
       ( \delta^{k\ell} - h^{k\ell} ) E_\ell,
\eea
and similarly for $H_{ab}$ and $H^*_{ab}$ with ($E_i$, $B_i$) replaced by ($D_i$, $H_i$).

Maxwell's equations in the normal frame in a general curved spacetime are derived in Eqs.\ (27)-(30) of \cite{Hwang-Noh-2023-EM-NL}. For gravitational waves in a flat background, assuming TT gauge and ignoring the sources, we have
\bea
   & & [ ( \delta^{ij} - h^{ij} ) D_j ]_{,i} = 0, \quad
       [ ( \delta^{ij} - h^{ij} ) D_j ]_{,0}
       - \eta^{ijk} \nabla_j H_k = 0,
   \nonumber \\
   & & [ ( \delta^{ij} - h^{ij} ) B_j ]_{,i} = 0, \quad
       [ ( \delta^{ij} - h^{ij} ) B_j ]_{,0}
       + \eta^{ijk} \nabla_j E_k = 0.
\eea
From Eq.\ (\ref{Hab-EH}), we have
\bea
   & & D_i = E_i - 2 \alpha \varrho \big[ B^2 - E^2
       - h^{jk} (B_j B_k - E_j E_k ) \big] E_i
       + 7 \beta \varrho (E^j B_j
       - h^{jk} E_j B_k ) B_i,
   \nonumber \\
   & & H_i = B_i - 2 \alpha \varrho \big[ B^2 - E^2
       - h^{jk} (B_j B_k - E_j E_k ) \big] B_i
       - 7 \beta \varrho (E^j B_j
       - h^{jk} E_j B_k ) E_i.
   \label{constitutive}
\eea
Setting $B_i \rightarrow B_i + b_i$, $H_i \rightarrow H_i + h_i$, $E_i \rightarrow e_i$, and $D_i \rightarrow d_i$, we have
\bea
   & & d^i_{\;,i} = 0,\quad
       d^i_{\;,0}
       - \eta^{ijk} \nabla_j h_k = 0, \quad
       b^i_{\;,i} = ( h^{ij} B_j )_{,i}, \quad
       b^i_{\;,0}
       + \eta^{ijk} \nabla_j e_k = ( h^{ij} B_j )_{,0},
   \label{ME-pert}
\eea
and
\bea
   d_i = ( 1 - 2 \alpha \varrho B^2 ) e_i
       + 7 \beta \varrho {\bf B} \cdot {\bf e} B_i, \quad
       h_i = ( 1 - 2 \alpha \varrho B^2 ) b_i
       - 2 \alpha \varrho ( 2 {\bf B} \cdot {\bf b}
       - h^{jk} B_j B_k ) B_i.
   \label{d-h}
\eea
{\it Assuming} a uniform and constant background magnetic field, we can derive
\bea
   & & ( \partial_0^2 - \Delta ) b_i
       = h_{i,00}^j B_j
       + 2 \alpha \varrho [
       ( 2 b_{k,\ell i}
       - h^{j}_{k,\ell i} B_j ) B^k B^\ell
       - ( 2 \Delta b_j
       - \Delta h_j^k B_k ) B^j B_i ]
       + 7 \beta \varrho \eta_i^{\;\;jk}
       e_{\ell, j 0} B^\ell B_k,
   \label{b-eq} \\
   & & ( \partial_0^2 - \Delta ) e_i
       = \eta_i^{\;\;jk} h_{k,j0}^\ell B_\ell
       - 2 \alpha \varrho
       \eta_i^{\;\;jk} ( 2 b_{\ell,j0}
       - h_{\ell,j0}^m B_m ) B^\ell B^k
       + 7 \beta \varrho
       ( e_{j,ki} B^k
       - e_{j,00} B_i ) B^j.
   \label{e-eq}
\eea

We align $z$-axis as the direction of gravitational wave propagation with $h_{ij} \propto e^{i \omega_g (x^0 - z)/c}$, and align the background magnetic field in the $y$-$z$ plane, thus $B_2 = B \sin{\theta}$, $B_3 = B \cos{\theta}$, and $B_1 = 0$. 
We have
\bea
   & & ( \partial_0^2 - \Delta ) b_1
       = h_{\times,00} B_2
       + 4 \alpha \varrho
       \partial_1 [ {\bf B} \cdot \nabla
       ( {\bf B} \cdot {\bf b} ) ]
       + 7 \beta \varrho \partial_0
       ( B_3 \partial_2 - B_2 \partial_3 )
       ( {\bf B} \cdot {\bf e} ),
   \label{b1-eq} \\
   & & ( \partial_0^2 - \Delta ) b_2
       = - h_{+,00} B_2 ( 1 + 2 \alpha \varrho B_2^2 )
       + 4 \alpha \varrho
       [ B_3 \partial_2 \partial_3
       - B_2 ( \partial_1^2 + \partial_3^2 ) ]
       ( {\bf B} \cdot {\bf b} )
       - 7 \beta \varrho
       B_3 \partial_0 \partial_1 ( {\bf B} \cdot {\bf e} ),
   \label{b2-eq} \\
   & & ( \partial_0^2 - \Delta ) b_3
       = 4 \alpha \varrho
       [ B_2 \partial_2 \partial_3
       - B_3 ( \partial_1^2 + \partial_2^2 ) ]
       ( {\bf B} \cdot {\bf b} )
       + 7 \beta \varrho
       B_2 \partial_0 \partial_1 ( {\bf B} \cdot {\bf e} ),
   \label{b3-eq}
\eea
and
\bea
   & & ( \partial_0^2 - \Delta ) e_1
       = h_{+,z0} B_2 ( 1 + 2 \alpha \varrho B_2^2 )
       - 4 \alpha \varrho \partial_0
       ( B_3 \partial_2 - B_2 \partial_3 )
       ( {\bf B} \cdot {\bf b} )
       + 7 \beta \varrho \partial_1
       [ {\bf B} \cdot \nabla
       ( {\bf B} \cdot {\bf e} ) ],
   \label{e1-eq} \\
   & & ( \partial_0^2 - \Delta ) e_2
       = h_{\times,z0} B_2
       + 4 \alpha \varrho
       B_3 \partial_0 \partial_1 ( {\bf B} \cdot {\bf b} )
       + 7 \beta \varrho (
       {\bf B} \cdot \nabla \partial_2
       - B_2 \partial_0^2 )
       ( {\bf B} \cdot {\bf e} ),
   \label{e2-eq} \\
   & & ( \partial_0^2 - \Delta ) e_3
       = - 4 \alpha \varrho
       B_2 \partial_0 \partial_1 ( {\bf B} \cdot {\bf b} )
       + 7 \beta \varrho (
       {\bf B} \cdot \nabla \partial_3
       - B_3 \partial_0^2 )
       ( {\bf B} \cdot {\bf e} ).
   \label{e3-eq}
\eea
Equations (\ref{GW-B-1}), (\ref{GW-B-2}) and (\ref{b1-eq})-(\ref{e3-eq}) provide a set of the graviton-photon conversion equations using the EM fields.

\subsubsection{In terms of the potential}

We have
\bea
   & & F^{0i} = \partial^i A_0 - \partial_0 A^i, \quad
       F^{ij} = \eta^{ijk}
       ( \delta_{k\ell} - h_{k\ell} ) B^\ell
       + \partial^i A^j - \partial^j A^i,
   \nonumber \\
   & &
       F^{*0i} = B^i + \eta^{ijk} \partial_j A_k, \quad
       F^{*ij} = \eta^{ijk} ( \partial_0 A_k - \partial_k A_0 ).
\eea
Taking the Coulomb gauge with $\nabla \cdot {\bf A} = 0$, and {\it assuming} a uniform and constant background magnetic field, Eq.\ (\ref{Fab-eq}) gives
\bea
   & & \Delta A_0 = 7 \beta \varrho B^i B^j \partial_i
       \partial_0 A_j,
   \label{A0-eq} \\
   & & ( \partial_0^2 - \Delta ) A_i
       = A_{0,i0}
       - \eta_i^{\;\;jk} B_\ell h^\ell_{k,j}
       + 2 \alpha \varrho \eta_i^{\;\;jk} B_k B_\ell
       [ 2 \eta^{\ell m n} \partial_m A_{n,j}
       + h^\ell_{m,j} B^m ]
       - 7 \beta \varrho A_{j,00} B^j B^i.
   \label{Ai-eq}
\eea
As we have $\Delta A_0 = 0$ to the $(\varrho B^2)^0$-th order, we set $A_0 = 0$ to that order. Using the potential, the homogeneous Maxwell's equation in (\ref{ME-Fab}) is identically valid.

Align $z$-axis as the gravitational wave propagation direction, and without losing generality, align the background magnetic field in $y$-$z$ plane, thus $B_2 = B \sin{\theta}$, $B_3 = B \cos{\theta}$, and $B_1 = 0$. We have
\bea
   & & \Delta A_0
       = 7 \beta \varrho {\bf B} \cdot \nabla
       \partial_0 ( {\bf B} \cdot {\bf A} ),
   \label{A0-eq-2} \\
   & & ( \partial_0^2 - \Delta ) A_1
       = \partial_0 \partial_1 A_0
       - B_2 h_{+,z} ( 1 - 2 \alpha \varrho B_2^2 )
       - 4 \alpha \varrho ( B_2 \partial_3 - B_3 \partial_2 )
       [ ( B_2 \partial_3 - B_3 \partial_2 ) A_1
       + \partial_1 ( B_3 A_2 - B_2 A_3 ) ],
   \label{A1-eq} \\
   & & ( \partial_0^2 - \Delta ) A_2
       = \partial_0 \partial_2 A_0
       - B_2 h_{\times,z}
       - 4 \alpha \varrho B_3 \partial_1
       [ ( B_2 \partial_3 - B_3 \partial_2 ) A_1
       + \partial_1 ( B_3 A_2 - B_2 A_3 ) ]
       - 7 \beta \varrho B_2 \partial_0^2
       ( {\bf B} \cdot {\bf A} ),
   \label{A2-eq} \\
   & & ( \partial_0^2 - \Delta ) A_3
       = \partial_0 \partial_3 A_0
       + 4 \alpha \varrho B_2 \partial_1
       [ ( B_2 \partial_3 - B_3 \partial_2 ) A_1
       + \partial_1 ( B_3 A_2 - B_2 A_3 ) ]
       - 7 \beta \varrho B_3 \partial_0^2
       ( {\bf B} \cdot {\bf A} ).
   \label{A3-eq}
\eea
Equations (\ref{GW-A-1}), (\ref{GW-A-2}) and (\ref{A0-eq-2})-(\ref{A3-eq}) provide a set of the graviton-photon conversion equations using the potential. Using Eq.\ (\ref{b-A}), we can show that Eqs.\ (\ref{b-eq}) and (\ref{e-eq}) are consistent with Eqs.\ (\ref{A0-eq}) and (\ref{Ai-eq}).

\section{Photon birefringence}
                                 \label{sec:birefringence}

Birefringence is a phenomena where the index of refraction $n$ depends on the polarization state of light. In {\it flat} spacetime, ignoring the gravitational waves, Eqs.\ (\ref{b-eq}) and (\ref{e-eq}) become
\bea
   & & ( \partial_0^2 - \Delta ) b_i
       = 4 \alpha \varrho B^j ( b_{j,ik} B^k
       - \Delta b_j B_i )
       + 7 \beta \varrho \eta_i^{\;\;jk} B^\ell e_{\ell,j0} B_k,
   \\
   & & ( \partial_0^2 - \Delta ) e_i
       = - 4 \alpha \varrho \eta_i^{\;\;jk}
       B^\ell b_{\ell,j0} B_k
       + 7 \beta \varrho B^j ( e_{j,ik} B^k
       - e_{j,00} B_i ).
\eea
In terms of the potential, in flat spacetime, Eqs.\ (\ref{A0-eq}) and (\ref{Ai-eq}) become
\bea
   & & \Delta A_0 = 7 \beta \varrho B^i B^j \partial_i
       ( \partial_0 A_j - \partial_j A_0 ),
   \label{A0-eq-flat} \\
   & & ( \partial_0^2 - \Delta ) A_i
       = A_{0,i0}
       + 4 \alpha \varrho \eta_i^{\;\;jk} B_k B_\ell
       \eta^{\ell m n} \partial_m A_{n,j}
       - 7 \beta \varrho A_{j,00} B^j B^i.
   \label{Ai-eq-flat}
\eea

\end{widetext}
We consider plane electromagnetic waves propagating in ${\bf k}$ direction, thus $e_i$, $b_i$, $A_i \propto e^{i (\omega t - {\bf k} \cdot {\bf x})}$. Align the $z$-axis as the background magnetic field direction, and without losing generality, align the ${\bf k}$ in $y$-$z$ plane with $k_2 = k \sin{\theta}$ and $k_3 = k \cos{\theta}$. The index of refraction is defined as $n \equiv kc/\omega$. From the above equations, using Eqs.\ (\ref{ME-pert}) and (\ref{d-h}), we can show
\bea
   n_\parallel
       = 1 + {7 \over 2} \beta \varrho B^2 \sin^2{\theta}, \quad
       n_\perp
       = 1 + 2 \alpha \varrho B^2 \sin^2{\theta},
\eea
where $n_\parallel$ is for $e_2$, $e_3$, $b_1$, $A_2$, and $A_3$ (i.e., electric field in the $y$-$z$ plane and magnetic field perpendicular to the plane), and $n_\perp$ for $e_1$, $b_2$, $b_3$, and $A_1$ (i.e., electric field perpendicular to the $y$-$z$ plane and magnetic field in the plane); it is essential to properly consider Eq.\ (\ref{A0-eq-flat}) where $A_0$ in the right-hand side disappears. This is the well-known birefringence due to the Euler-K\"ockel-Heisenberg corrections \cite{Toll-1952, Erber-1961, Klein-1964, Adler-1971, Brezin-1971, Greiner-Reinhardt-2009}.

%
%
%
\section{Discussion}
                                         \label{sec:Discussion}

The nonlinear corrections in electrodynamics like the Euler-K\"ockel-Heisenberg type or a more general form in Eq.\ (\ref{L}) can be interpreted as an effective medium property as presented in Sec.\ \ref{sec:NL}; see Eq.\ (\ref{constitutive}) for the constitutive relation \cite{Euler-Kockel-1935}, now including the gravitons.

Similarly as the nonlinear correction causes birefringence in photon propagation, it also causes the parity breaking effect in the graviton propagation. We note that only $\alpha$-term affects the graviton conversion equation, whereas both $\alpha$ and $\beta$-terms affect the photon conversion equation as well as the photon birefringence.

We showed that the two methods of handling photons give apparently different equations. As one set of equations can be derived from the other using the relation between the potential and EM fields, these two methods coincide in the full equations level. However, if we focus on part of equations, like the graviton propagation equations ignoring the photon part, the difference demands attention; for example, see Eq.\ (\ref{k-EM}) compared with Eq.\ (\ref{k-A}). In such a case, we suggest to consider the equations in terms of the EM fields as the physical one to compare with experiment, naturally so as the EM fields are the ones associated with measurement by an observer. In this sense, in the presence of gravity, the calculation based the potential should be properly translated to the ones using the EM fields.

The Euler-K\"ockel-Heisenberg correction is valid in the low energy range of QED, with the photon energy lower than the electron mass energy. Evaluating in Gaussian unit, we have $\varrho \sim 2.65 \times 10^{-24} / ({\rm Tesla})^2$. The nonlinear terms are one-loop corrections to classical electrodynamics and are valid for $\hbar \omega \ll m_e c^2$ \cite{Euler-Kockel-1935} and $B < B_c$ \cite{Heisenberg-Euler-1936} with the critical field strength introduced as $B_c \equiv m_e^2 c^3/(e \hbar) = 4.41 \times 10^9 {\rm Tesla}$ in Gaussian unit. Thus, $\varrho = {\bar \alpha \over 45 \pi}B_c^{-2}$, and in QED context, the above nonlinear correction has amplitude $\varrho B^2 = 5.16 \times 10^{-5} (B/B_c)^2$.

Due to extreme small value of the coupling $\varrho$ in QED case, experimental measurement of the nonlinear corrections in the graviton is not expected in foreseeable future. Despite the small coupling, experimental efforts are made to measure the photon birefringence and other consequences due to the nonlinear correction in a strong magnetic field without gravity, for reviews see \cite{Ejlli-2020, Fedotov-2023}.

%
%
%
\section*{Acknowledgments}

We thank Dr.\ Seokhoon Yun for useful discussion. We wish to thank two reviewers for their corrections and helpful suggestions. H.N.\ was supported by the National Research Foundation (NRF) of Korea funded by the Korean Government (No.\ 2018R1A2B6002466 and No.\ 2021R1F1A1045515). J.H.\ was supported by IBS under the project code, IBS-R018-D1, and by the NRF of Korea funded by the Korean Government (No.\ NRF-2019R1A2C1003031).

%
%


\end{document}